\documentclass[10pt]{article}
\begin{document}

\begin{center}
{\large \bf Thermal Expansion Puzzles}
\end{center}
\vspace{0.1in}

\begin{center}

{Rajesh R. Parwani\footnote{Email: parwani@nus.edu.sg}}

\vspace{0.3in}

{Department of Physics and\\}
{University Scholars Programme,\\}
{National University of Singapore,\\}
{Kent Ridge, Singapore.}

\vspace{0.3in}
\end{center}
\vspace{0.1in}
\begin{abstract}
The standard formula that describes the thermal expansion of a solid  creates several puzzles for discerning students. Three puzzles are reviewed, and their common resolution discussed both conceptually and quantitatively.

\end{abstract}

\vspace{0.5in}

\section{Which starting length?}
Undergraduate textbooks provide the following formula for use in calculating the change in length, in some direction, of a solid heated between two temperatures $T_o$ and $T_f$,
\begin{equation}
L_f  = L_o( 1 + \alpha \Delta T) \label{basic}
\end{equation} 
with $\Delta T \equiv T_f - T_o$ and $\alpha$ the coefficient of linear thermal expansion. Some books do mention that the formula is approximate, valid when $\alpha$ is small, and that experimentally $\alpha$ is constant for small 
$\Delta T$ \cite{Serway}.  

However even with those cautions, puzzles arise. The first: the  change in length, $\Delta L = L_f - L_o$, in (\ref{basic}) depends on the starting length $L_o$; but if the length is continually changing with  tempertaure, then how should the formula be used? For example, one can imagine the same physical process but performed in $N$ small steps involving temperature changes $\delta T \equiv \Delta T /N$, so that the length at the end of step $n$ is given by   
\begin{eqnarray}
L_n = L_{n-1} (1 + \alpha \delta T) \, , 
\end{eqnarray}
assuming, for simplicity of discussion, that $\alpha$ is a constant independent of temperature. After $N$ steps one achieves a temperature change of $\Delta T$, and the length becomes 
\begin{equation}
L_N = L_o (1 + \alpha \delta T)^N \, . \label{insteps}
\end{equation}
This expression is clearly very different from (\ref{basic}) eventhough the physical process, a temperature change of $\Delta T$,  is identical! For $\alpha \delta T$ small one may  truncate the binomial expansion of (\ref{insteps}), giving 
\begin{equation}
L_N \approx L_o (1 + \alpha \Delta T)\, ,
\end{equation}   
which does agrees with (\ref{basic}). Through this example, {\it one concludes that formula (\ref{basic}) is self-consistent only to order  $\alpha \Delta T$}. 

Of course it would be nice to have a formula that is consistent to all orders. The way to obtain it is to write (\ref{basic}) as 
\begin{equation}
\Delta L = \alpha L \Delta T \, ,
\end{equation} 
which emphasizes that the change in $L$ depends on the value of $L$ at that moment. So the proper procedure is to consider infinitesimal changes and write the process in calculus notation,
\begin{equation}
{d L \over dT} = \alpha L \, ,
\end{equation} 
an equation that can be easily integrated to give \cite{bart1}
\begin{equation}
L = L_o \exp[A(T)] \, , \label{exact}
\end{equation}
where $A(T) \equiv \int_{T_o}^{T} \alpha dT $. The expression (\ref{exact}) is valid even when $\alpha$ is not a constant and can also be used for large $\alpha$. 

As discussed in Ref.\cite{bart1}, a virtue of the exact equation (\ref{exact}) is that it indicates the two approximations required to obtain (\ref{basic}): firstly, for $\alpha$ approximately constant, one may write $A(T) \approx \alpha \Delta T$. Then a power series expansion of the exponential gives the usual textbook formula, correct to order $\alpha \Delta T$, in agreement with the conclusion reached above using self-consistency arguments.

The result (\ref{exact}) can be obtained in a slightly different way which emphasizes the expansion of the solid in small steps: If $\alpha$ changes with temperature, then the generalisation of (\ref{insteps}) is 
 \begin{eqnarray}
L_N &=& L_o \Pi_{i=1}^{i=N} (1 + \alpha_i \delta T) \, \label{insteps2}
\end{eqnarray}
where $\alpha_i$ is the coefficient of expansion during the small interval $i$.
Thus 
\begin{eqnarray}
\ln {L_N \over L_o} &=& \sum_{i=1}^{i=N} \ln (1 + \alpha_i \delta T) \\
&=& \sum_{i=1}^{i=N} \alpha_i \delta T + O(1/N) ,  \\
\end{eqnarray}
where a Taylor series expansion of the logarithm was used. Then taking the limit $N \to \infty$, while keeping $\Delta T = N \delta T$ fixed gives, using the definition of an integral, 
\begin{equation}
\ln {L \over L_o} = \int_{T_o}^{T} \alpha dT \, ,
\end{equation}
which is the same as Eq.(\ref{exact}).

\section{Irreversibility?}
A puzzle that has been much discussed in the literature is the apparent irreversibility of length change as predicted by (\ref{basic}). For, applying successive changes of $\Delta T$ and $-\Delta T$ leads to 
\begin{eqnarray}
L_f &=& L_o (1 + \alpha \Delta T) (1 - \alpha \Delta T) \\
&=& L_o ( 1 - (\alpha \Delta T)^2 ) \label{shrink}
\end{eqnarray} 
and a shrinking rod \cite{shrink}. However, there is clearly no shrinkage when the exact expression (\ref{exact}) is used, as noted in \cite{bart2}. 

That the result (\ref{shrink}) is doubtful could  have been deduced  as follows. As discussed in the last section, the expression (\ref{basic}) is valid only to order $\alpha \Delta T$ and so any results obtained by using it cannot be trusted at subleading orders. In particular, the negative quadratic term in (\ref{shrink}) is unreliable. So again, this puzzle evaporates when one uses consistency arguments.

\section{When can a  rod pass through a ring?}

Suppose that at a temperature $T_o$ rod with circular cross-section has a diameter $d_1$ while a ring has an inner diameter of $d_2 < d_1$. The coefficient of linear expansion of the rod is $\alpha_1$ while that of the ring $\alpha_2$. To simplify the discussion, let us assume that the expansion coefficients are temperature independent. Is it possible to make the rod pass through the ring by heating them to some common higher temperature? Intuitively one would expect that the answer is yes as long as the ring expands faster than the rod, $\alpha_2 > \alpha_1$. However a direct calculation using (\ref{basic}) produces an unexpected conclusion. Let us investigate this.

After a temperature change of $\Delta T$ the relevant diameters become,
\begin{eqnarray}
d_{2f} &=& d_2 (1 + \alpha_2 \Delta T) \, , \\
d_{1f} &=& d_1 (1+ \alpha_1 \Delta T) \, .
\end{eqnarray}

For the rod to pass through the ring requires $d_{2f} \ge d_{1f}$, which when applied to the previous two equations gives
\begin{equation}
(\alpha_2 d_2 - \alpha_1 d_1) \Delta T \ge (d_1 - d_2) \, .
\end{equation}

Since $d_1 > d_2$ and since $\Delta T$ has been assumed to be positive, there is a physical solution if and only if,
\begin{equation}
\alpha_2 d_2 - \alpha_1 d_1 > 0 \, , \label{cond1}
\end{equation}   

giving
\begin{equation}
\Delta T \ge {(d_1 - d_2) \over (\alpha_2 d_2 - \alpha_1 d_1) } \, . \label{tapprox}
\end{equation}

The constraint (\ref{cond1}) is physically obscure. It may be written as $d_2 / d_1 > \alpha _1 / \alpha_2$ and when combined with $d_1 > d_2$ it  implies the expected condition
\begin{equation}
\alpha_2 >  \alpha_1  \, . \label{cond2}
\end{equation}  
However, as the derivation above shows, the desired outcome (\ref{cond2}) is only a necessary condition but not sufficient by itself, the stronger condition (\ref{cond1}) being required. By now the reader must have guessed that the conclusion (\ref{cond1}) might be an artifact of using the approximate relation (\ref{basic}). That is indeed the case: Performing a similar analyis but adopting instead the exact expression (\ref{exact}) with constant expansion coefficients gives
\begin{equation}
{d_{2f} \over d_{1f} } = {d_2 \over d_1} \exp[(\alpha_2 - \alpha_1) \Delta T] \, .
\end{equation}
For the rod to pass through the ring, the right-hand-side of the last equation must be at least equal to one, and taking logarithms gives
\begin{equation}
 (\alpha_2 - \alpha_1) \Delta T \ge \ln {d_1 \over d_2} > 0 \, ,
\end{equation}
which is satisfied once $\alpha_2 > \alpha_1$. That is, an exact analysis does show the intuitive condition to be {\it sufficient}.

It is instructive to compare  the exact minimum temperature change at which the rod passes through the ring, 
\begin{equation}
\Delta T_{ex} = {1 \over (\alpha_2 - \alpha_1)} \ln  {d_1 \over d_2}  \,  , \label{texact}
\end{equation}
with the approximate expression from (\ref{tapprox}). The approximate expression should follow from the exact value when the relative expansions are small, meaning $\alpha \Delta T \ll 1$. In that regime, the diameters of the rod  and ring must be very close, $ d_1 - d_2 \sim O (\alpha \Delta T)$. So
\begin{eqnarray}
\ln{ d_1 \over d_2}  &=& \ln { (d_1 -d_2) + d_2 \over d_2 } \\
&=& \ln \left( 1 + { (d_1 -d_2) \over d_2} \right) \\
&\approx &  { (d_1 -d_2) \over d_2}  \, 
\end{eqnarray}
to leading order. Then (\ref{texact}) can be written 
\begin{eqnarray}
\Delta T_{ex} & \approx  & {(d_1 - d_2) \over (\alpha_2 d_2 - \alpha_1 d_2) } \\
& \approx  & {(d_1 - d_2) \over (\alpha_2 d_2 - \alpha_1 d_1)}
\end{eqnarray}
where in the denominator of the last line one has again approximated $d_1$ by $d_2$ in one term. This derived expression agrees with the approximate result that follows from (\ref{tapprox}). 
\newpage

\section{Summary}
Assuming for simplicity that $\alpha$ is temperature independent, a self-consistency argument was used to show that the standard textbook formula (\ref{basic}) could only be correct to order $\alpha  \Delta T$.
This is equivalent to saying that only small relative expansions can be dealt with using that formula, a condition which is fortunately satisfied in common applications because of the tiny value of $\alpha$, in common units, for real solids. 

The various paradoxes result from using the standard formula beyond its regime of validity.  For more precise investigations one should use the exact expression (\ref{exact}) which can deal not only with large expansions but also includes the more realistic case of a temperature varying $\alpha$. The expression (\ref{exact}) was originally derived in Ref.\cite{bart1} and has been re-derived here through an explicit limiting procedure.


\begin{thebibliography}{99}

\bibitem{Serway} R.A. Serway and J.W. Jewett. Jr., Physics for Scientists and Engineers (Thomson-Brooks/Cole, Sixth Edition).  

\bibitem{bart1} R.A. Bartels, Am. J. Phys. 41 (1973) 78. 

\bibitem{shrink} T. P. Toepker, Am. J. Phys. 55 (1987) 177;  \\
F.C. Stephenson, Am. J. Phys. 55 (1987) 777; \\
R.C. Good, Jr.,  Am. J. Phys. 55 (1987) 971.  

\bibitem{bart2} R.A. Bartels, Am. J. Phys. 56 (1988) 570. 




\end{thebibliography}
\end{document}